\documentclass{ws-ijmpb}
\usepackage{color}

\begin{document}
\markboth{Victor Lakhno}{(s-f(d) exchange mechanism of magnon generation by slow spinpolarons)}
\title{s-f(d) exchange mechanism of magnon generation by slow spinpolarons}
\author{Victor Lakhno}
\address{Russian Academy of Sciences, Institute of Mathematical Problems of Biology,\\Pushchino, Moscow Region, 142290, Russia \\
lak@impb.psn.ru}
\maketitle

\begin {abstract}
It is shown that in a single-axis antiferromagnetic semiconductor
placed in a strong magnetic field, dispersionless magnons start emitting at any arbitrarily small velocity of an electron occurring in a spinpolaron state. If magnons are dispersed they are generated when the spinpolaron velocity exceeds the minimum phase velocity of magnons. The maximum power of magnon generation caused by the drift of spinpolarons is estimated.
\end {abstract}


\section{ Introduction}

The possibility to construct a generator of magnons on the basis of electron-magnon interaction has been discussed for nearly 50 years\cite{lit1}$^-$\cite{lit3}. Interest in this problem is associated with the possibility of designing a submillimeter-range laser possessing unique characteristics which would emit at a frequency controllable by the magnetic field.

The idea that a beam of fast electrons can excite spin waves (SW) in magnetically-ordered crystals was first suggested in Ref. \refcite{lit4}. There an assumption was made that excitation of SW occurs as a result of relativistic interaction of the beam with the spins of magnetic atoms. Since this interaction has the value $\sim V^2/c^2$ where $V$ is the beam velocity, $c$ is the speed of light, then the effect may be realized at the velocities close to the speed  of light. In subsequent papers, in order to enhance the interaction of electrons with the magnetic moments of the atoms, an idea was put forward to realize this effect in magnetic semiconductors where the concentrations of charge carriers is  several orders of magnitude larger than the concentration which can be realized in an electron beam in vacuum.

However, practical attempts to realize generation of SW with the use
of semiconductor plasma of ferromagnetic semiconductors have failed
(results of some of them are presented in Ref.~\refcite{lit5} and in
more recent publications\cite{lit6}$^-$\cite{lit8}).

As was shown in Ref.~\refcite{lit9,lit10} we get qualitatively new possibilities for creation of a magnon generator if we use not ferro-, but antiferromagnetic  (AF) semiconductors. In this case generation of magnons is possible due to a strong s-f(d) exchange interaction not involving a relativistic factor.
In Refs.~\refcite{lit1,lit3,lit9,lit10} it was assumed that spin waves are amplified by charge carriers in delocalized states of band type.
Here the main problem hindering the effect realization is associated with the necessity to get a drift velocity of current carriers  higher than the phase velocity of SW $V_{\phi}$. However, in overwhelming majority of antiferromagnetics, the current carriers have low velocity and very strong electric fields are needed to speed them up so that their velocity should exceed
$V_{\phi}$.

In this paper we suggest a qualitatively different principle of magnon generation by current carriers. It can be realized when charge carriers are in localized spinpolaron states. In this case the generation takes place with the use of s-f(d) exchange interaction which does not require that the current carriers should obey the condition $V > V_{\phi}$.
The states of small-radius spinpolaron type occur in narrow-band antiferromagnetic semiconductors: $W_b\ll AS/2$, where $W_b$  is the band width, $A$ is s-f(d) exchange integral, $S$ is the  spin of a magnetic atom.\cite{lit11} Their effective mass is several orders of magnitude exceeds that of a free electron and therefore they are slightly mobile if at all. The drift velocity of such charge carriers will be small in an electric field, therefore their contribution into generation of magnons will be small too.

For this reason it will be interesting to consider another type of spinpolarons, i.e. large-radius ones which occur in wide-band antiferromagnetics. As is shown in Ref.~\refcite{lit11}, in the absence of an external magnetic field, charge carriers can be self-localized only in antiferromagnetics with extremely low Neel temperatures. In Ref.~\refcite{lit11} such spinpolaron formations were termed ferron states. Ferron states, similarly to spinpolaron states in narrow-band antiferromagnetics are slightly mobile, therefore their drift velocity cannot be large. In rather a strong external magnetic field, which however, does not exceed the value of collapse field, ferron states disappear.
As was shown in Ref.~\refcite{lit12}, quite a different situation is observed in antiferromagnetics with high Neel temperatures.
Most antiferromagnetics are high-temperature with sublattices collapse field $H_c\ge 10^6$ Oe and self-localized ferron states cannot form there. If, however, such an antiferromagnetic is placed in a quantizing magnetic field, sinpolaron states become energetically more advantageous.\cite{lit12} Being large-radius states, they can move in an electric field. As will be shown below, motion of such spinpolarons leads to emission of magnons and can be used for constructing a magnon generator.

\section{Hamiltonian of a single axis antiferromagnetic.}

In the general form, the Hamiltonian describing the motion of a conductivity electron in an anisotropic antiferromagnetic placed in an electric field looks like:\cite{litA1}

\begin{eqnarray}
\label{eq.1v}
{\cal {H}} = {\cal {H}}_e + {\cal {H}}_{int} + {\cal {H}}_M\\ \nonumber
{\cal {H}}_e = \frac {1}{2m} (\vec {P} + \frac {e}{c}\vec {A})^2, \qquad
{\cal {H}}_{int} = -\sum _{m,m'} A(R_m - R_{m'})(S_{m'}, \sigma _{m'}),
\\ \nonumber
{\cal {H}} _M = -\frac {1}{2} \sum _{i, m_1\ne m_1'} I^i_{m_1m_1'} S^i_{m_1}S^i_{m_1'} -\frac {1}{2} \sum _{i, m_2\ne m_2'} I^i_{m_2m_2'} S^i_{m_2}S^i_{m_2'}\\ \nonumber
- \sum _{i, m_1, m_2} I^i_{m_1m_2} S^i_{m_1}S^i_{m_2}-\sum _{i, m_1} S^i_{m_1}H_i
- \sum _{i, m_2} S^i_{m_2}H_i
\end{eqnarray}
where ${\cal {H}}_e$ describes the electron motion in a magnetic field with the vector potential $\vec A$, ${\cal {H}}_{int}$ is responsible for the interaction between the conductivity electron and the magnetic  subsystem of the crystal with the exchange constant s - f (d) which hereafter we will put equal to $A_{mm`} = A \delta_{mm`}$, ${\cal {H}}_M$ is an exchange Hamiltonian of an anisotropic antiferromagnetic  AF placed in a magnetic field  $H$ ($H$ is given in energy units). For a single axis antiferromagnetic whose anisotropy axis lies along $Y$-axis, exchange integrals have the form:

\begin{eqnarray}
\label{eq.2v}
I^x_{mm'} = I^z_{mm'},\qquad I^y_{mm'} = I_{mm'} + \Delta I_{mm'}
\end{eqnarray}
Restricting our consideration to the case when interaction takes place only between two equivalent sublattices, we introduce the notation:

\begin{eqnarray}
\label{eq.3v}
\sum _{m_1,m_2} I_{m_1,m_2}S^2 = NJ_{12},\qquad
\sum _{m_1,m_2} \Delta I_{m_1,m_2}S^2 = N \Delta J_{12},
\end{eqnarray}
where $N$ is the number of magnet atoms in the sublattice.

Let us consider the case when the external magnetic field $\vec H$ is directed along the axis $Z$, that is perpendicular to the anisotropy axis. In this case over the whole interval of $H$ variation, the vector of the antiferromagnetic magnetization  is directed along the field $(\Delta J_{12} < 0)$. In the spin-wave approximation, spin operators are replaced by the Bose operators of the birth and annihilation of spin deviations and in their own  coordinate system (associated with the direction of the magnetic atom spin) are given by:

\begin{eqnarray}
\label{eq.4v}
S^z_m = S - b^+_mb_m, \qquad  S^+_m = \sqrt {2S} b_m,\qquad
S^-_m = \sqrt {2S} b^+_m,
\end{eqnarray}
where $S^{\pm}_m = S^x_m \pm iS^y_m$. Believing the electron spin to be fully polarized, i.e. having put $\sigma _m = \sigma ^z \delta (R_m - r)$ and taking into account short-range character of the s -f exchange interaction $A(R_m - R_{m`}) = A\delta (R_m - R_{m`})$, we express the interaction Hamiltonian ${\cal {H}}_{int}$ with the use of Eq.~(\ref{eq.4v}) as:

\begin{eqnarray}
\label{eq.5v}
{\cal {H}}_{int} = \frac {A}{2}S^z_r,\qquad S^z_r=\sum _m S^z_m\delta (r - R_m),\\ \nonumber
S^z_m = (S - b^+_mb_m)\cos \Theta _m + \sqrt{\frac {S}{2}}\sin \Theta _m(b_m + b^+_m)
\end{eqnarray}
The quantization axis here is the direction of the magnetic field (general coordinate system; $\Theta _m$ is the angle between the spin direction $S_m$ and the external magnetic field). Turning in Eq.~(\ref{eq.1v}), Eq.~(\ref{eq.5v}) from spin deviation operators Eq.~(\ref{eq.4v}) to representation of the operators of the birth and annihilation of magnons $\xi ^+$, $\xi $ with the use of canonical Bogolyubov-Tyablikov transformation:

\begin{eqnarray}
\label{eq.6v}
b_m=\sum _q(U_{mq}\beta _q + U^*_{mq}\beta ^+_q),
\end{eqnarray}
which makes the quadratic form of  ${\cal {H}}_{M}$ diagonal, we will get the Hamiltonian of an electron-magnon system with an interaction linear in magnon operators $\beta _q$ in the form:\cite{lit11}

\begin{eqnarray}
\label{eq.1}
\hat H =  \frac {1}{2m}(\vec P + \frac {e}{c}\vec A)^2 + \frac {Q}{\sqrt N}\sum _k (\beta _k\,e^{i\,k\,r} + c.c.)
+ \sum _k \hbar \, \omega _M \,\beta ^+_k \,\beta _k,\\\nonumber
Q = \frac {A}{4} \,\frac {(SH_{EA})^{1/2}(1 - H^2/H^2_E)^{3/4}}
{[H_E(1 - H^2/H^2_E) + 2H_A]^{1/2}},\\\nonumber
\hbar \,\omega _M = \mu \,H^{}_{EA}(1 - H^2/H^2_E),\quad
H_{EA} = \sqrt {H_EH_A}
\end{eqnarray}
In Eq.~(\ref{eq.1}) $H_E$ is the exchange field of AF sublattices
collapse; $H_A = 2|\Delta J_{12}|$, is the  field of magnetic
anisotropy of a single axis AF; $N=N_xN_yN_z$ is the number of atoms
in the crystal; $\mu = 2\mu _B$, where $\mu _B = (e\hbar)/(2m_0c)$
Bohr magneton. Notice that, according to Ref.~\refcite{lit13}, the
part of the interaction Hamiltonian quadratic in magnon operators,
except for the limit case $H\to H_E$, makes a much less
contribution, than its linear part Eq.~(\ref{eq.1}).

\section{ Holstein Hamiltonian for AF in a quantizing magnetic field}

In rather a strong magnetic field with $\hbar\omega _c > \hbar\omega _M $, $\Delta E$ where $\Delta E$ is a change in the electron energy caused by interaction with magnons, $\omega _c=eH/mc$ is a cyclotron frequency, the radial part of the wave function of Hamiltonian Eq.~(\ref{eq.1}) has the form:
\begin{eqnarray}
\label{eq.2}
R(\rho)=\frac {1}{\sqrt{2\pi\rho^2_0}}\exp \{{-\rho ^2/4\rho ^2_0}\}
\end{eqnarray}
where $\rho _0 = \sqrt{2\mu _B\ c\ \hbar/eH}$ is a magnetic length, $\mu _B$ is Bohr magneton. With the use of Eq.~(\ref{eq.2}) Hamiltonian Eq.~(\ref{eq.1}) takes the form:
\begin{eqnarray}
\label{eq.3}
\hat {\tilde H} = \langle R|\hat H|R\rangle = \frac {\hbar \,\omega _c}{2} -
\frac {\hbar ^2}{2m} \,\frac {\partial ^2}{\partial z^2} + \sum _k \hbar \,\omega _M \,\beta ^+_k\, \beta _k  +\\\nonumber
\frac {Q}{\sqrt N}\sum _k (\beta _k\, e^{-k^2_{\perp}\,\rho ^2_0}\,e^{i\,k_zz} + c.c.).
\end{eqnarray}
In what follows we will reckon the energy from the quantity $\hbar \omega _c/2$ omitting this term in the expression for Hamiltonian. Quantities $k_z$ and $k_{\perp}$ involved in Eq.~(\ref{eq.3}) are longitudinal and perpendicular components of the wave vector $\vec k$.

Using the well-known procedure\cite{lit14} we can reduce Hamiltonian Eq.~(\ref{eq.3}) to effective one-dimensional Hamiltonian by canonical transformation of magnon operators $\beta _k$ to operators $a_{k_{z,s}}$:
\begin{eqnarray}
\label{eq.4}
a_{k_z,s} = \sum _{k_{\perp}} \beta _kf_s(k_{\perp}, k_z),\quad
\beta _k = \sum _sa_{k_{z,s}}f^*_s(k_{\perp}, k_z),
\end{eqnarray}
where $f_s(k_{\perp}, k_z)$ is the total system of functions satisfying the relations:
\begin{eqnarray}
\label{eq.5}
\sum _sf_s(k_{\perp}, k_z) f^*_s(k^{\prime}_{\perp}, k_z) = \delta _{ k_{\perp}, k'_{\perp}},\\\nonumber
\sum _{k_{\perp}} f^*_s(k_{\perp}, k_z) f_{s'}(k_{\perp}, k_z) = \delta _{ s,s'},\quad
s = 0, 1, 2, \ldots
\end{eqnarray}
In the case of a quantizing magnetic field considered here we choose:
\begin{eqnarray}
\label{eq.6}
f_0=\sqrt{\frac {4\,\pi \, \rho ^2_0}{S_0}}\,e^{-k^2_{\perp} \, \rho ^2_0/2}
\end{eqnarray}
where $S_0=N_xa_xN_ya_y$; $a_x$, $a_y$ are lattice constants in $x$ and $y$ directions.

The use of Eq.~(\ref{eq.4})~--~Eq.~(\ref{eq.6}) transforms Hamiltonian $\hat{\tilde H}$ to the form:
\begin{eqnarray}
\label{eq.7}
\hat {\tilde H} =
\sum _{k_z,\, s\ne 0} \hbar \,\omega _M \,a ^+_{k_{zs}}\, a _{k_{zs}}
+
\sum _{k_z} \hbar \,\omega _M \,a ^+_{k_z}\, a _{k_z} \\ \nonumber
- \frac {\hbar ^2}{2m}  \, \frac {\partial ^2}{\partial z^2}
+ \frac {Q}{\sqrt N}\sum _{k_z}\sqrt {\frac {S_0}{4 \, \pi \, \rho ^2_0}}
(a_{k_z}\, e^{i\,k_z\,z} + c.c.),
\end{eqnarray}
where we put $a_{k_z} = a_{k_z,0}$.

Hamiltonian Eq.~(\ref{eq.7}) exactly coincides with Holstein Hamiltonian in structure\cite{lit15,lit16} up to a non-essential renormalization of total magnon field energy due to the presence of magnon modes non-interacting with the particle.

In the next section we present a brief report of the results for a large-radius spinpolaron in the limit of weak and strong coupling in the case of Hamiltonian
 Eq.~(\ref{eq.7}).

\section{ Spinpolaron in the limit of weak and strong coupling}

The spinpolaron ground state energy $E(k)$ in the case of weak interaction of an electron with a magnetic subsystem in the second order of the perturbation theory is given by the expression:\cite{lit12}
\begin{eqnarray}
\label{eq.a1}
E(k) = \frac{\hbar ^2k^2}{2m} + \frac{1}{N_z}\sum _{k_z} \frac{Q^2_0}{\varepsilon _{k_z} - \varepsilon _{k_z-q} - \hbar\omega_M - i0},\\\nonumber
\varepsilon_{k_z} = \hbar ^2k^2_z/2m, \quad
Q_0 = Q\sqrt {a_xa_y/4\pi \rho ^2_0}
\end{eqnarray}
which yields the shift of the electron energy with respect to the bottom of the conductivity band $\Delta E $ equal to:
\begin{eqnarray}
\label{eq.a2}
\Delta E = -\alpha \hbar \omega _M,\quad
\alpha = Q^2_0\sqrt{ma^2/2\hbar^2(\hbar\omega _M)^3}
\end{eqnarray}
In the strong coupling limit the electron energy $W$ is determined by Schr\"odinger equation:
\begin{eqnarray}
\label{eq.a3}
\left(-\frac{\hbar ^2}{2m}\Delta _z + U(z) - W\right)\varphi _0(z) = 0
\end{eqnarray}
\begin{eqnarray}
\label{eq.a4}
U(z) = \frac{1}{N_z\hbar}\sum _{k_z}\frac{2Q^2_0\omega _M}{(\omega _M)^2 -  (Vk_z)^2}
\cdot \int e^{-ik_zz^{\prime}}|\varphi (z^{\prime})|^2dz^{\prime}e^{ik_zz}
\end{eqnarray}
where $V$ - is the velocity of spinpolaron; $\varphi _0$ - is the wave function of spinpolaron ground state.
For $V=0$, the solution of spectrum problem Eq.~(\ref{eq.a3}), Eq.~(\ref{eq.a4}) yields:
\begin{eqnarray}
\label{eq.a6}
\varphi _0(z) = \frac{1}{\sqrt {2r_z}}\mbox{ch} ^{-1}\frac {z}{r_z}, \quad
W = - \frac {\hbar ^2}{2mr^2_z},
\quad  r_z = \frac{\hbar ^2}{ma}\frac {\hbar\omega _M}{Q^2_0},
\end{eqnarray}
From Eq.~(\ref{eq.a2}) and Eq.~(\ref{eq.a4}) it follows that the spinpolaron energy  in the strong coupling limit ($E=W/3$) can be written in the form:
\begin{eqnarray}
\label{eq.a8}
E = - \alpha ^2\hbar \omega_M/3
\end{eqnarray}
Comparison of Eq.~(\ref{eq.a2}) with Eq.~(\ref{eq.a8}) demonstrates that if $\alpha < 3 $ the case of weak coupling is realized with delocalized electron states of Bloch type. This case was considered in Ref.~\refcite{lit13}. For the case $\alpha > 3 $ which will be considered below the strong coupling limit is determinated by electron wave function given by Eq.~(\ref{eq.a6}).

\section{ Equilibrium velocity of a spinpolaron in an electric field, power of magnon emission}

The problem of equilibrium velocity of a large-radius Holstein polaron is considered in detail in papers Ref.~\refcite{lit17,lit18}. In the absence of dissipation the number of magnons $N$ excited in a unit of time during the process of spinpolaron motion is equal to the rate at which the spinpolaron magnon "surrounding"  looses its energy $2 Im\,(W/\hbar)$:
\begin{eqnarray}
\label{eq.b1}
\frac {dN}{dt} = 2Im \frac {W}{\hbar}
\end{eqnarray}
\begin{eqnarray}
\label{eq.b2}
Im W = \frac {\pi}{N_z\hbar} \sum _k \frac {2\omega _MQ^2_0}{(\omega _M + kV)}
\left|\int e^{-ik_zz}|\varphi (z)|^2dz \right|^2
\cdot \delta (\omega _M - V\cdot k_z)
\end{eqnarray}
where $ImW$ is the imaginary part of energies Eq.~(\ref{eq.a3}) and Eq.~(\ref{eq.a4}) of the moving electron.
To obtain Eq.~(\ref{eq.b2}) we have used the equality:
$$
\frac {1}{\omega _M - k_zV + i 0} = P \frac {1}{\omega _M - k_zV}
- i\pi \delta (\omega _M - k_zV) .
$$
When electron moves in an external electric field $\mathcal {E}$, the spinpolaron equilibrium velocity is determined from the balance of the energies which a spinpolaron gets from the field $\mathcal {E}$ and gives to excite the magnon subsystem:
\begin{eqnarray}
\label{eq.b3}
\hbar \,\omega _M \frac {dN}{dt} = e\,\mathcal {E}\,V = 2\,\omega _M\,Im\,W
\end{eqnarray}
With the use of Eq.~(\ref{eq.b1}), Eq.~(\ref{eq.b2}), Eq.~(\ref{eq.b3}) the dependence of the equilibrium velocity $V$ on $\mathcal {E}$ takes the form:
\begin{eqnarray}
\label{eq.b4}
e\,\mathcal {E}\,V = \frac {\pi ^2}{4}\,\frac {Q^2_0\;r^2_z\;a\;\hbar\;\omega_M^3}{V^3}
{\mbox{sh}^{-2}\left(\frac{\pi}{2}\,\frac{r_z\;\omega _M}{V}\right)}
\end{eqnarray}
The typical form of $V(\mathcal {E})$ dependence is shown in FIG.1.
Relation Eq.~(\ref{eq.b4}) suggests that the maximum possible stationary velocity of a polaron $V_{max}$ caused by emission of magnons is equat to:
\begin{eqnarray}
\label{eq.b5}
V_{max} \cong \frac {\pi}{4}\,r_z\,\omega _M
\end{eqnarray}
Then the maximum power of magnon emission $P_{max}$ generated by electrons in AF with concentration $n$ in the unit volume will be:
\begin{eqnarray}
\label{eq.b6}
P_{max} = e\,\mathcal {E}_{max}\,V_{max}\,n
\end{eqnarray}
From Eq.~(\ref{eq.b4}) - Eq.~(\ref{eq.b6}) for $P_{max}$  we will get:
\begin{eqnarray}
\label{eq.b7}
P_{max} \approx \frac {16}{\pi\;\mbox{sh}^2\;2}\,
\frac {Q^2_0\;a\;n}{\hbar\;r_z}
\end{eqnarray}

\begin{figure}
\includegraphics[scale=0.7]{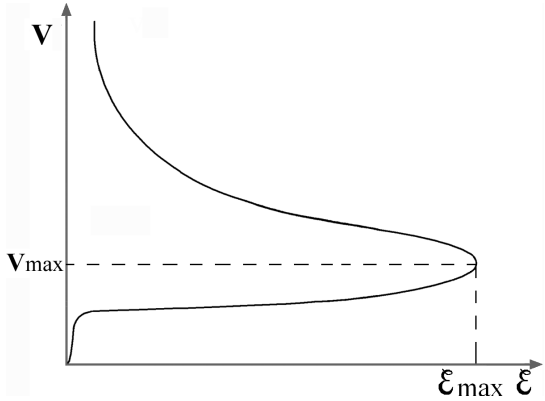}
\caption{The dependence of the particle velocity V on electric field $\mathcal {E}$
in the strong coupling limit}
\label{f1}
\end{figure}

\section{ Assessment of the effect and discussion of the results}

The mechanism considered in the paper resolves the problem of magnon generation in AF. As distinct from a coherent mechanism of amplification of spin waves considered in Ref.~\refcite{lit13}, in the case investigated here, generation of nonequilibrium magnons occurs in a non-coherent way. The fact that magnons demonstrate attenuation does not alter the result: it is only the time of existence of generated nonequilibrium magnons which is changed.

In paper Ref.~\refcite{lit13} a theory of amplification of spin waves in a quantizing magnetic field was constructed for non-localized electron states. Such states are formed if the condition of weak coupling is met when a shift of the electron energy $\Delta E$ determined by Eq.~(\ref{eq.a1}) exceeds the spinpolaron energy, i.e. when the inequality $\alpha < 3$ is fulfilled. For the parameter values used in Ref.~\refcite{lit13}, this inequality is fulfilled, i.e. spinpolaron states are not formed and estimates obtained in Ref.~\refcite{lit13} are valid for them.

To illustrate when spinpolaron states can be formed let us take the parameter values to be: $A=1$ eV, $S=1$, $a=3 {\AA}$, $(a_x=a_y=a_z=a)$, $H_E=10^6$ Oe,
$H_{EA} = 5\cdot 10^4$ Oe,
 $H = 9\cdot 10^5$ Oe, $m=m_0$. This corresponds to magnon frequency $\hbar \omega _M \approx  10^{-4}$ eV. The chosen values of $H_E$ and $H_{EA}$ are typical for a wide range of AF for example for manganese compounds such as MnX, $X=F, Se, O, Te, Ge, Hg$ etc.

In this case the value $\alpha \approx 4$, and the condition of strong coupling necessary for the formation of spinpolarons is fulfilled.
The maximum power of magnon generation, by Eq.~(\ref{eq.b7}), for $n=10^{17}$ cm$^{-3}$, will be: $P_{max} \cong  5\cdot 10^6$W, at $V_{max}\approx 7\cdot 10^4$cm/sec.

Taking account of the dispersion of the magnon frequency can change considerably the results obtained. According to Ref.~\refcite{lit19}, for a single axis AF, consideration of the dependence of the magnon frequency $\omega _M(k)$ on the wave vector $k$ is given by the expression:
\begin{eqnarray}
\label{eq.14}
\omega ^2_M(k) = \omega ^2_M + V^2_0k^2,\quad V_0 = \frac {2 \, \mu \,B}{\hbar \, M_0} \sqrt {1 - H^2/H^2_E }
\end{eqnarray}
where $M_0$ is the magnetic moment of AF sublattice, $B$ is an exchange constant whose characteristic value is equal to $B\approx H_EM_0/a^2$.
 As is shown in Ref.~\refcite{lit17,lit18}, for dispersion law Eq.~(\ref{eq.14}), the emission of magnons arises only when $V > V_0$. For the parameter values presented above, Eq.~(\ref{eq.14}) leads to $V_0 \approx 1,3\cdot 10^3 \sqrt {1 - H^2/H^2_E}$ cm/sec.
It should be stressed that condition $V > V_0$ does not place any restrictions on the value of the wave vector of spin waves $k$, in contrast to Cherenkov mechanism, when the condition $V\,k > \omega _M$ leads to generation of ultra-high frequency magnons $k > 10^5$ cm$^{-1}$ for $V < V_{\max}$ in which case magnons not only fail to be transformed into electromagnetic radiation, but they cannot be generated or even exist either.

Finally, we note that we have obtained maximum possible estimate of the power of the magnon generator. Obviously, the actual power can be some orders of magnitude less.

Nevertheless, realization of the effect considered holds out a hope that the power obtained per a unit volume will be some orders of magnitude larger  than the power of a free-electron-laser, which is the only device capable of emitting submillimeter- controllable radiation.

By virtue of the fact that constant quantizing magnetic field is practically difficult to achieve, the effect can easier be realized in the impulse synchzonized magnetic and electric fields.

The work is supported by RFBR project N09-07-12073; N10-07-00112.


\end{document}